\documentclass[12pt,preprint]{aastex}

\newcommand{\dechms}[4]{$#1^{\rm h}#2^{\rm m}#3\mbox{$^{\rm s}\mskip-7.6mu.\,$}#4$}

\newcommand{\decdms}[4]{$#1^{\circ}#2'#3\mbox{$''\mskip-7.6mu.\,$}#4$}
\newcommand{\msec}[2]{$#1\mbox{$''\mskip-7.6mu.\,$}#2$}
\newcommand{\mmsec}[2]{$#1\mbox{$^s\mskip-7.6mu.\,$}#2$}

\newcommand{\Lsun}{L$_{\odot}$}
\newcommand{\Msun}{M$_{\odot}$}
\newcommand{\Rsun}{R$_{\odot}$}

\begin{document}

\title{VLBA determination of the distance to nearby star-forming regions\\
       II. Hubble 4 and HDE~283572 in Taurus}

\author{Rosa M. \ Torres, Laurent Loinard}
\affil{Centro de Radiostronom\'{\i}a y Astrof\'{\i}sica, 
       Universidad Nacional Aut\'onoma de M\'exico,\\
       Apartado Postal 72--3 (Xangari), 58089 Morelia, Michoac\'an, M\'exico;\\
       r.torres@astrosmo.unam.mx}

\author{Amy J.\ Mioduszewski}
\affil{National
    Radio Astronomy Observatory, Array Operations Center,\\ 1003
    Lopezville Road, Socorro, NM 87801, USA}

\and

\author{Luis F.\ Rodr\'{\i}guez}
\affil{Centro de Radiostronom\'{\i}a y Astrof\'{\i}sica,
       Universidad Nacional Aut\'onoma de M\'exico,\\
       Apartado Postal 72--3 (Xangari), 58089 Morelia, Michoac\'an, M\'exico}

\begin{abstract} 

The non-thermal 3.6 cm radio continuum emission from the naked T Tauri
stars Hubble 4 and HDE~283572 in Taurus has been observed with the
Very Long Baseline Array (VLBA) at 6 epochs between September 2004 and
December 2005 with a typical separation between successive
observations of 3 months. Thanks to the remarkably accurate astrometry
delivered by the VLBA, the trajectory described by both stars on the
plane of the sky could be traced very precisely, and modeled as the
superposition of their trigonometric parallax and uniform proper
motion. The best fits yield distances to Hubble 4 and HDE~283572 of
132.8 $\pm$ 0.5 and 128.5 $\pm$ 0.6 pc, respectively. Combining these
results with the other two existing VLBI distance determinations
in Taurus, we estimate the mean distance to the Taurus association
to be 137 pc with a dispersion (most probably reflecting the depth of
the complex) of about 20 pc.
\end{abstract}


\section{Introduction} 

While our understanding of the main sequence evolution of Solar-type
stars is now very solid, our comprehension of their youth is
significantly less advanced. Increasingly detailed pre-main sequence
theoretical models, as well as improved observational constraints are
clearly needed, and are actively sought (see Klein et al.\ 2007, White
et al.\ 2007, G\"udel et al.\ 2007 for recent reviews). On the
observational front, significant progress is currently being made
thanks to large on-going X-ray and infrared surveys of nearby
star-forming regions carried out with space observatories (e.g.\ Evans
et al.\ 2003, G\"udel et al.\ 2007). It should be noticed, however,
that some of the stellar parameters (luminosity, mass, etc.) most
relevant to constrain theoretical models depend critically both on the
quality of the data used to estimate them {\bf and} on the distance to
the object under consideration. Unfortunately, the distance to even
the nearest star-forming complexes (e.g.\ Taurus or $\rho-$Ophiuchus)
are not known to better than 20\% (Elias 1978a,b, Kenyon et al.\ 1994,
Knude \& Hog 1998, Bertout \& Genova 2006). This is, in part, a
consequence of the fact that the otherwise highly successful Hipparcos
mission (Perryman et al.\ 1997) performed comparatively poorly in
star-forming regions (Bertout et al.\ 1999) because young stars
--being still heavily embedded in their parental clouds-- are faint in
the optical bands observed by Hipparcos.

\clearpage
\begin{deluxetable}{llllllr}
\rotate
\tablewidth{0pt}
\tablecaption{Measured source positions and fluxes}
\tablehead{
\colhead{Mean UT date}    &  
\colhead{$\alpha$ (J2000.0)} &
\colhead{$\sigma_\alpha$} &
\colhead{$\delta$ (J2000.0)} &
\colhead{$\sigma_\delta$} &
\colhead{$F_\nu$} &
\colhead{$\sigma$}\\%
~~(yyyy.mm.dd ~~ hh:mm)~~ & & & & & (mJy) & ($\mu$Jy)}
\startdata
Hubble 4:\\%
2004.09.19 ~~ 11:47 \dotfill & \dechms{04}{18}{47}{0327419} & \mmsec{0}{0000020} & \decdms{28}{20}{07}{398977} & \msec{0}{000050} & 0.67 & 54 \\%
2005.01.04 ~~ 04:46 \dotfill & \dechms{04}{18}{47}{0319609} & \mmsec{0}{0000022} & \decdms{28}{20}{07}{389009} & \msec{0}{000071} & 0.76 & 73 \\%
2005.03.25 ~~ 23:44 \dotfill & \dechms{04}{18}{47}{0318775} & \mmsec{0}{0000009} & \decdms{28}{20}{07}{381391} & \msec{0}{000021} & 4.66 & 114 \\%
2005.07.04 ~~ 16:51 \dotfill & \dechms{04}{18}{47}{0328115} & \mmsec{0}{0000022} & \decdms{28}{20}{07}{375000} & \msec{0}{000053} & 0.65 & 58 \\%
2005.09.18 ~~ 11:52 \dotfill & \dechms{04}{18}{47}{0330740} & \mmsec{0}{0000019} & \decdms{28}{20}{07}{370321} & \msec{0}{000040} & 1.25 & 53 \\%
2005.12.28 ~~ 05:15 \dotfill & \dechms{04}{18}{47}{0323418} & \mmsec{0}{0000012} & \decdms{28}{20}{07}{360573} & \msec{0}{000025} & 1.53 & 51 \\%
\hline
\\[-0.15cm]%
HDE~283572:\\%
2004.09.22 ~~ 11:35 \dotfill & \dechms{04}{21}{58}{8521561} & \mmsec{0}{0000004} & \decdms{28}{18}{06}{389421} & \msec{0}{000010} & 7.13 & 81 \\%
2005.01.06 ~~ 04:39 \dotfill & \dechms{04}{21}{58}{8514573} & \mmsec{0}{0000048} & \decdms{28}{18}{06}{380015} & \msec{0}{000091} & 0.92 & 58 \\%
2005.03.30 ~~ 23:34 \dotfill & \dechms{04}{21}{58}{8514676} & \mmsec{0}{0000022} & \decdms{28}{18}{06}{372534} & \msec{0}{000038} & 1.71 & 65 \\%
2005.06.23 ~~ 17:34 \dotfill & \dechms{04}{21}{58}{8523648} & \mmsec{0}{0000007} & \decdms{28}{18}{06}{367852} & \msec{0}{000014} & 4.23 & 80 \\%
2005.09.23 ~~ 11:32 \dotfill & \dechms{04}{21}{58}{8528216} & \mmsec{0}{0000070} & \decdms{28}{18}{06}{363175} & \msec{0}{000140} & 0.52 & 62 \\%
2005.12.24 ~~ 05:31 \dotfill & \dechms{04}{21}{58}{8522172} & \mmsec{0}{0000028} & \decdms{28}{18}{06}{354808} & \msec{0}{000070} & 0.51 & 47
\enddata
\end{deluxetable}
\clearpage

Future space missions such as GAIA will certainly be able to detect
stars much fainter than those accessible to Hipparcos, but these
missions will still be unable to access the most deeply embedded
populations, and are still at least a decade away. Radio observations
with Very Long Baseline Interferometers (VLBI) provide an interesting
alternative avenue, because they can deliver extremely accurate
absolute astrometry (better than 0.1 mas) if proper calibration is
applied. In the last few years, such observations have proven capable
of measuring the trigonometric parallax of sources within a few
kiloparsecs of the Sun with a precision of a few percents (Brisken et
al.\ 2000, 2002, Loinard et al.\ 2005, 2007, Xu et al.\ 2005,
Hachisuka et al. 2006, Hirota et al.\ 2007, Sandstrom et al.\
2007). Because the sensitivity of VLBI experiments is limited, only
compact non-thermal emitters can usually be detected. In star-forming
regions, two kinds of such non-thermal sources exist: masers and
magnetically active young stars. Masers are ubiquitous in regions
where massive stars are formed, but they are absent or unpredictably
variable in low- and intermediate mass star-forming sites. Low-mass
young stars, on the other hand, tend to have active magnetospheres
that can generate detectable non-thermal continuum emission (e.g.\
Andr\'e et al.\ 1992, Feigelson \& Montmerle 1999, Dulk 1985). Thus,
the distance to nearby star-forming regions can be measured very
accurately if adequate non-thermal sources are identified in them, and
multi-epoch observations are obtained over the course of a few
years. This method has been successfully applied to water and methanol
masers in nearby massive star-forming regions (Xu et al.\ 2005,
Hachisuka et al. 2006, Hirota et al.\ 2007) and to the non-thermal
continuum emission associated with low-mass T Tauri stars (Loinard et
al.\ 2005, 2007, Sandstrom et al.\ 2007). In all these cases, a
precision typically an order of magnitude better than previous
estimates was achieved. Since adequate non-thermal sources are
available in essentially all the nearby sites of star formation,
multi-epoch VLBI observations have the potential of improving
significantly our knowledge of the space distribution of star-forming
regions around the Sun. With this goal in mind, we have initiated a
large project aimed at accurately measuring the trigonometric parallax
of a sample of magnetically active young stars in the most prominent
and often-studied northern star-forming regions within 1 kpc of the
Sun (Taurus, $\rho-$Ophiuchus, Perseus, Serpens, and Cepheus; the
distance to Orion has already been measured using VLBI
techniques--Hirota et al.\ 2007, Sandstrom et al.\ 2007) using the
10-element Very Long Baseline Array (VLBA) of the National Radio
Astronomy Observatory (NRAO). In the present article, we will
concentrate on HDE~283572 and Hubble 4, two young stars in
Taurus. This will allow us to examine in more detail the distribution
and kinematics of young stars in this important star-forming region.

\clearpage
\begin{deluxetable}{llllllr}
\tablecaption{Julian dates and Earth coordinates for Hubble 4 and HDE~283572}
\tablehead{
\colhead{Mean UT date}    &
\colhead{JD}      &
\multicolumn{3}{c}{Earth Barycentric coordinates} \\%
(yyyy.mm.dd ~~ hh.mm) & & \multicolumn{3}{c}{Astronomical Units}}
\startdata
Hubble4:\\%
2004.09.19 ~~ 11:47 \dotfill & 2453267.99  &   $+$1.006998486 &  $-$0.052084106 &  $-$0.022682627 \\% 
2005.01.04 ~~ 04:46 \dotfill & 2453374.70  &   $-$0.231331103 &  $+$0.875675935 &  $+$0.379526720 \\%
2005.03.25 ~~ 23:44 \dotfill & 2453454.48  &   $-$0.990029933 &  $-$0.069134209 &  $-$0.030092055 \\%
2005.07.04 ~~ 16:51 \dotfill & 2453556.20  &   $+$0.228244142 &  $-$0.908947748 &  $-$0.394190784 \\%
2005.09.18 ~~ 11:52 \dotfill & 2453631.99  &   $+$1.005815421 &  $-$0.069425778 &  $-$0.030225422 \\%
2005.12.28 ~~ 05:15 \dotfill & 2453732.72  &   $-$0.107253794 &  $+$0.898552365 &  $+$0.389430205 \\%
\hline
\\[-0.15cm]%
HDE~283572:\\%
2004.09.22 ~~ 11:35 \dotfill & 2453270.98 & $+$1.007690418 &  $-$0.005037832 &  $-$0.002285078 \\% 
2005.01.06 ~~ 04:39 \dotfill & 2453376.69 & $-$0.265058725 &  $+$0.867496343 &  $+$0.375981760 \\% 
2005.03.30 ~~ 23:34 \dotfill & 2453460.46 & $-$0.978622564 &  $-$0.163101149 &  $-$0.070826870 \\% 
2005.06.23 ~~ 17:34 \dotfill & 2453545.23 & $+$0.044465499 &  $-$0.930888246 &  $-$0.403696737 \\% 
2005.09.23 ~~ 11:32 \dotfill & 2453636.98 & $+$1.007380117 &  $+$0.008930420 &  $+$0.003741574 \\% 
2005.12.24 ~~ 05:31 \dotfill & 2453728.73 & $-$0.037702136 &  $+$0.903636825 &  $+$0.391632177 \\% 
\enddata
\end{deluxetable}
\clearpage

Hubble 4 is a K7 naked T Tauri star with an effective temperature of
4060 K (Brice\~no et al.\ 2002). It has long been known to have a
particularly active magnetosphere that produces non-thermal radio
emission characterized by significant variability, large circular
polarization and a nearly flat spectral index (Skinner 1993). It was
detected in VLBI experiments, with a flux of a few mJy by Phillips et
al.\ (1991), and is also an X-ray source (G\"udel et al.\ 2007). The
superficial magnetic field of Hubble 4 has been estimated to be about
2.5 kG using Zeeman-sensitive Ti I lines (Johns-Krull et al.\
2004). HDE~283572, on the other hand, is a somewhat hotter
($T_{\mbox{eff}} = $ 5770 K --Kenyon \& Hartmann 1995) G5 naked T
Tauri star. Early observations with the Einstein satellite showed that
it has a fairly bright X-ray counterpart (Walter et al.\ 1987). It was
initially detected as a radio source by O'Neal et al.\ (1990), and in
VLBI observations by Phillips et al.\ (1991) with a flux of about 1
mJy.

\section{Observations and data calibration}

In this paper, we will make use of a series of six continuum 3.6 cm
(8.42 GHz) observations of Hubble 4 and HDE~283572 obtained roughly
every three months between September 2004 and December 2005 with the
VLBA (Tab.\ 1). Our pointing directions were at $\alpha_{J2000.0}$ =
\dechms{04}{18}{47}{033}, $\delta_{J2000.0}$ =
+\decdms{28}{20}{07}{398}, and $\alpha_{J2000.0}$ =
\dechms{04}{21}{58}{847}, $\delta_{J2000.0}$ =
+\decdms{28}{18}{06}{502} for Hubble 4 and HDE~283572, respectively.
Each observation consisted of series of cycles with two minutes on
source, and one minute on the main phase-referencing quasar J0429+2724
(the same for both targets). Each 24 minutes, we also observed three
secondary calibrators (J0433+2905, J0408+3032, and J0403+2600) forming
a triangle around the astronomical source (Fig.\ 1). All four
calibrators are very compact extragalactic sources whose absolute
positions are known to better than 1 milli-arcsecond (Beasley et al.\
2002).

\clearpage
\begin{figure}[!b]
\centerline{\includegraphics[height=0.65\textwidth,angle=-90]{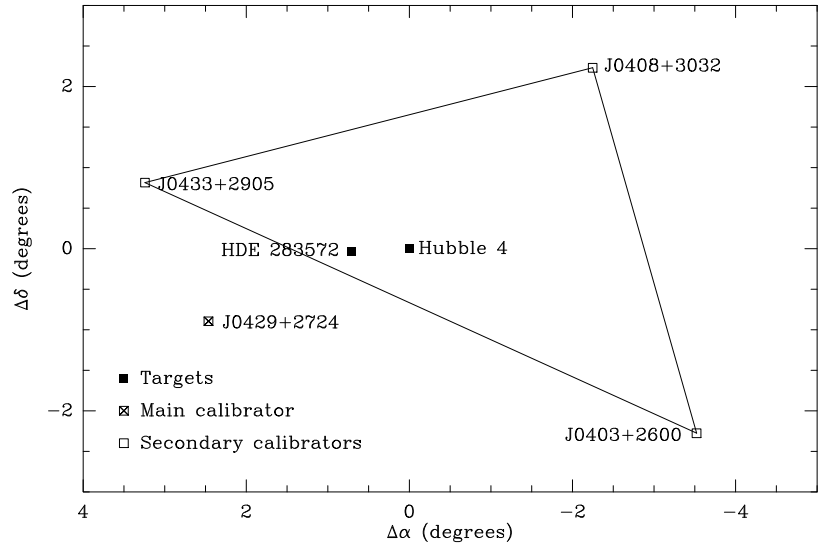}}
\caption{Relative position of the astronomical targets (Hubble 4 and 
HDE~283572), the main calibrator (J0429+2724), and the secondary
calibrators (J0433+2905, J0408+3032, and J0403+2600).}
\end{figure}
\clearpage

The data were edited and calibrated using the Astronomical Image
Processing System (AIPS --Greisen 2003). The basic data reduction
followed the standard VLBA procedures for phase-referenced
observations. First, the most accurate measured Earth Orientation
Parameters obtained from the US Naval Observatory database were
applied to the data to improve the values initially used by the VLBA
correlator.  Second, dispersive delays caused by free electrons in the
Earth's atmosphere were accounted for using estimates of the electron
content of the ionosphere derived from Global Positioning System (GPS)
measurements. {\it A priori} amplitude calibration based on the
measured system temperatures and standard gain curves was then
applied. The fourth step was to correct the phases for antenna
parallactic angle effects, and the fifth was to remove residual
instrumental delays caused by the VLBA electronics. This was done by
measuring the delays and phase residuals for each antenna and IF using
the fringes obtained on a strong calibrator. The final step of this
initial calibration was to remove global frequency- and time-dependent
phase errors using a global fringe fitting procedure on the main phase
calibrator (J0429+2724), which was assumed at this stage to be a point
source.

In this initial calibration, the solutions from the global fringe fit
were only applied to the main phase calibrator itself. The
corresponding calibrated visibilities were then imaged, and several
passes of self- calibration were performed to improve the overall
amplitude and phase calibration. In the image obtained after the
self-calibration iterations, the main phase calibrator is found to be
slightly extended.  To take this into account, the final global fringe
fitting part of the reduction was repeated using the image of the main
phase calibrator as a model instead of assuming it to be a point
source. Note that a different phase calibrator model was produced for
each epoch to account for possible small changes in the main
calibrator structure from epoch to epoch. The solutions obtained after
repeating this final step were edited for bad points and applied to
the astronomical targets and to the main and secondary calibrators.

The astrometry precision of VLBI observations such as those presented
here, depends critically on the quality of the phase
calibration. Systematic errors, unremoved by the standard calibration
procedures described above, usually dominate the phase calibration
error budget, and limit the astrometric precision achieved to several
times the value expected theoretically (e.g.\ Fomalont 1999, Pradel et
al.\ 2006). At the frequency of the present observations, the main
sources of systematic errors are inaccuracies in the troposphere model
used, as well as clock, antenna and {\it a priori} source position
errors. These effects combine to produce a systematic phase difference
between the calibrator and the target, causing position shifts. One
effective strategy to measure and correct these systematic errors
consists of including observations of more than one phase calibrator
chosen to surround the target (Fomalont \& Kogan 2005).  This allows
phase gradients around the source due to errors in the troposphere
model or related to uncertainties in the cataloged position of the
calibrators, to be measured and corrected. This strategy was applied
to our observations using the three secondary calibrators mentioned
earlier (Fig.\ 1), and resulted in significant improvements in the
final phase calibration and image quality.

Because of the time spent on the calibrators, only about 5 of the 9
hours of telescope time allocated to each of our observations were
actually spent on source. Once calibrated, the visibilities were
imaged with a pixel size of 50 $\mu$as after weights intermediate
between natural and uniform (ROBUST = 0 in AIPS) were applied. This
resulted in typical r.m.s.\ noise levels of 50--80$\mu$Jy beam$^{-1}$
(Tab.\ 1). Both sources were detected with a signal to noise ratio
better than 10 at each epoch (Tab.\ 1). The source absolute positions
at each epoch (also listed in Tab.\ 1) were determined using a 2D
Gaussian fitting procedure (task JMFIT in AIPS). This task provides an
estimate of the position error (columns 3 and 5 of Tab.\ 1) based on
the expected theoretical astrometric precision of an interferometer:

\begin{equation} 
\sigma = {\lambda \over 2 \pi B} {1 \over SNR}, \label{error} 
\end{equation}

\noindent where $\lambda$ is the wavelength, $B$ the baseline, and
$SNR$ the image signal-to-noise ratio (Thompson et al.\ 1986). In
spite of the extra calibration steps taken to improve the phase
calibration, uncorrected systematic errors still exist, and must be
added quadratically to the values deduced from Eq.\ 1. These remaining
systematic errors are difficult to estimate {\it a priori}, and may
depend on the structure of the source under consideration. Here, we
will estimate these systematic effects from the fits to the data (see
below).

\clearpage
\begin{figure*}
\centerline{\includegraphics[width=0.9\textwidth,angle=270]{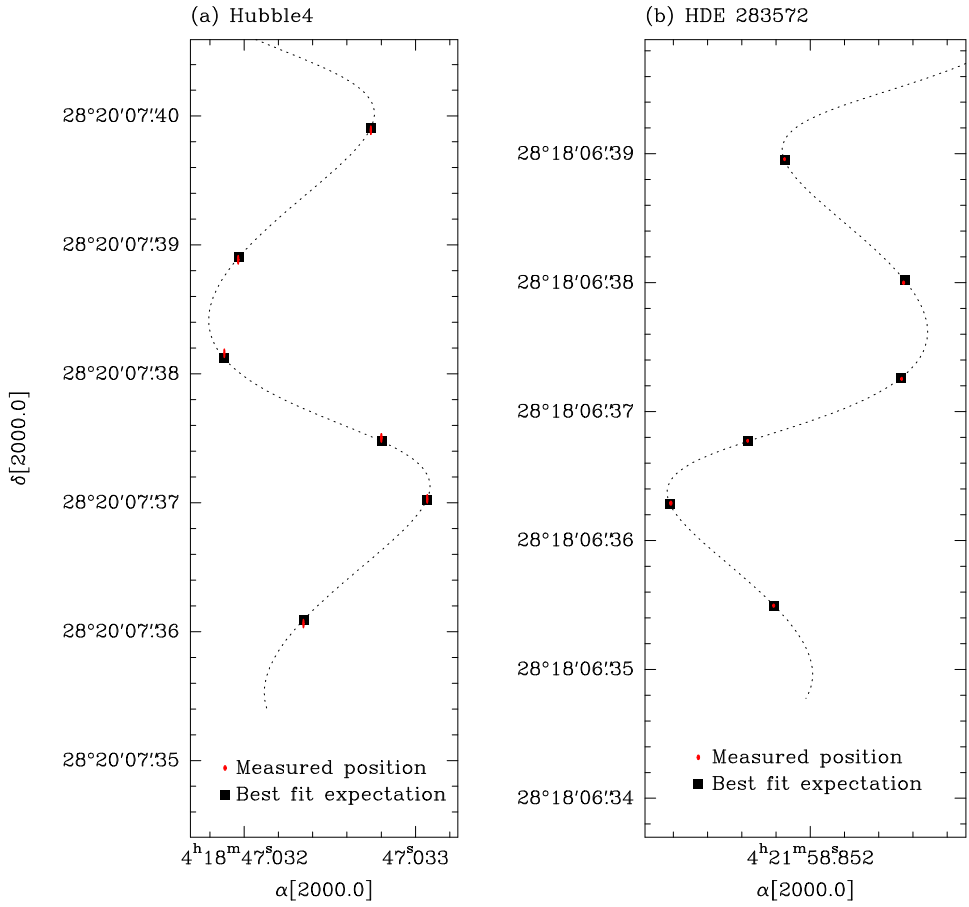}}
\caption{Measured positions and best fit for (a) Hubble 4, and (b)
HDE~283572. The observed positions are shown as ellipses, the size of
which represents the error bars.}  
\end{figure*}

\begin{figure*}
\centerline{\includegraphics[width=0.5\textwidth,angle=270]{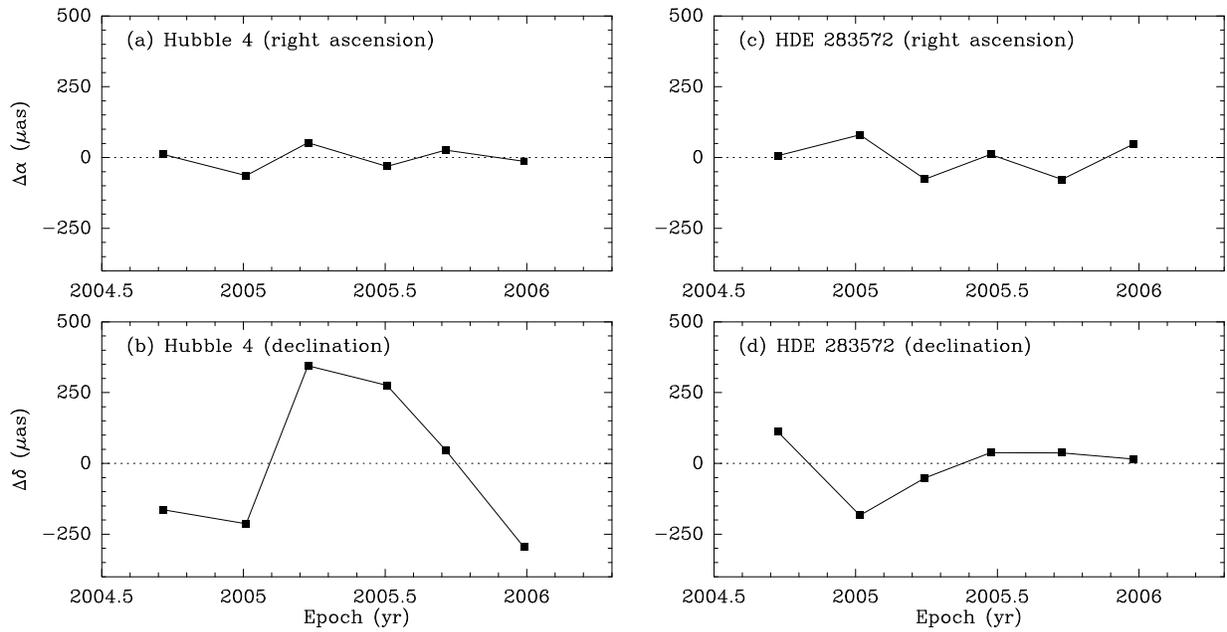}}
\caption{Post-fit residuals for Hubble 4 (left) and HDE~283572 (right)
in right ascension (top) and declination (bottom).}
\end{figure*}
\clearpage

\section{Results}

The displacement of the sources on the celestial sphere is the
combination of their trigonometric parallax ($\pi$) and proper motions
($\mu$). The reference epoch was taken at the mean epoch of each
series of observations: JD 24353500 ($\equiv$ 2005.355) for bothy
sources.  Since the sources considered here appear to be isolated, we
considered linear and uniform proper motions. The astrometric
parameters were calculated using the SVD-decomposition fitting scheme
described by Loinard et al.\ (2007). The necessary barycentric
coordinates of the Earth, as well as the Julian date of each
observation were calculated using the Multi-year Interactive Computer
Almanac (MICA) distributed as a CDROM by the US Naval
Observatory. They are given explicitly in Tab.\ 2 for all epochs and
sources. The best fits give the following parameters:

\begin{eqnarray}
\alpha_{J2005.355} & = & \mbox{ \dechms{04}{18}{47}{032414} } ~ \pm ~ \mbox{ \mmsec{0}{000001} } \nonumber \\%
\delta_{J2005.355} & = & \mbox{ \decdms{28}{20}{07}{3792} } ~ \pm ~ \mbox{ \msec{0}{0002} } \nonumber \\%
\mu_\alpha & = & 4.30 ~ \pm ~ 0.05 ~ \mbox{mas yr$^{-1}$} \nonumber \\%
\mu_\delta & = & -28.9 ~ \pm ~ 0.3 ~ \mbox{mas yr$^{-1}$} \nonumber \\%
\pi & = & 7.53 ~ \pm ~ 0.03 ~ \mbox{mas,} \nonumber
\end{eqnarray}

\noindent
and

\begin{eqnarray}
\alpha_{J2005.355} & = & \mbox{ \dechms{04}{21}{58}{852030} } ~ \pm ~ \mbox{ \mmsec{0}{00002} } \nonumber \\%
\delta_{J2005.355} & = & \mbox{ \decdms{28}{18}{06}{37128} } ~ \pm ~ \mbox{ \msec{0}{00005} } \nonumber \\%
\mu_\alpha & = & 8.88 ~ \pm ~ 0.06 ~ \mbox{mas yr$^{-1}$} \nonumber \\%
\mu_\delta & = & -26.6 ~ \pm ~ 0.1 ~ \mbox{mas yr$^{-1}$} \nonumber \\%
\pi & = & 7.78 ~ \pm ~ 0.04 ~ \mbox{mas,} \nonumber
\end{eqnarray}

\noindent for Hubble 4 and HDE~283572, respectively. The measured
parallaxes correspond to distances of 132.8 $\pm$ 0.5 pc for Hubble 4,
and 128.5 $\pm$ 0.6 pc for HDE~283572. The post-fit rms (dominated by
the remaining systematic errors mentioned at the end of Sect.\ 2) is
quite good for HDE~283572: 60 $\mu$as and 90 $\mu$as in right
ascension and declination, respectively. For Hubble 4, on the other
hand, the residual is good in right ascension (40 $\mu$as), but large
in declination (240 $\mu$as). To obtain a reduced $\chi^2$ of one both
in right ascension and declination, one must add quadratically 3.1
microseconds of time and 340 microseconds of arc to the formal errors
delivered by JMFIT for Hubble 4, and 4.3 microseconds of time and 115
microseconds of arc for HDE~283572. All the errors quoted in the paper
include these systematic contributions.

The origin of the large declination residual for Hubble 4 (which does
not affect strongly the parallax determination, because the latter is
dominated by the right ascension measurements) is not entirely
clear. The fact that the residual is only (or, at least, mostly)
detected in declination (Fig.\ 3) would suggest a calibration issue.
Indeed, astrometric fitting of phase-referenced VLBI observations is
usually worse in declination than in right ascension (e.g.\ Fig.\ 1 in
Chatterjee et al.\ 2004) as a result of residual zenith phase delay
errors (Reid et al.\ 1999). We consider this possibility fairly
unlikely here, however, because such a problem would have been
detected during the multi-source calibration, and because the
observations and reduction of Hubble 4 and HDE~283572 (which does not
appear to be affected by any calibration issue) were performed
following identical protocols and over the same period of
time. Another element that argues against a calibration problem is
that the large residual is not the result of one particularly
discrepant observation: in addition to the fit mentioned above where
all 6 observations of Hubble 4 are taken into account, we made 5 fits
where we sequentially discarded one of the epochs. All 5 fits gave
similar astrometric parameters, and a similarly large declination
residual. Thus, we argue that this large residual might be real,
rather than related to a calibration problem. At the distance of
Hubble 4, 240 $\mu$as correspond to 0.032 AU, or about 7 \Rsun. Hubble
4 is estimated to have a radius of about 3.4 \Rsun\ (Johns-Krull et
al.\ 2004), so the amplitude of the residual is just about 2
$R_*$. Baring this figure in mind, at least two mechanisms could
potentially explain the large declination residual: (i) the
magnetosphere of Hubble 4 could be somewhat more extended than its
photosphere, and the residuals could reflect variations in the
structure of the magnetosphere; (ii) Hubble 4 could have a companion,
and the residuals could reflect the corresponding reflex motion. Let
us examine the pros and cons of these two possibilities.  

If the residuals were related to a variable extended magnetosphere,
one would expect the emission to be occasionally somewhat extended.
Interestingly, Phillips et al.\ (1991) reported that Hubble 4 was
slightly resolved in their VLBI data, and we find it to be resolved
also in at least two of our own observations. On the other hand, if
the emission were related to variations in the magnetosphere, one
would expect to see variations with the periodicity of the rotational
period of the star (about 12/sin{\it i} days --Johns-Krull et al.\
2004). Given that the separation between our successive observations
is typically three months, we would expect the residuals to be
essentially random. Instead, those residuals seem to show a
periodicity of about 1.2 years (Fig.\ 3b). This would be more
consistent with our alternative proposal that the residuals be related
to the reflex motion of Hubble 4 due to the presence of an unseen
companion. The semi-major axis corresponding to a period of 1.2 yr and
a mass of 0.7 \Msun\ (see below) is just about 1 AU. Since the ratio
between the amplitude of the reflex motion and that of the orbital
path is the inverse of the ratio between the mass of the primary and
that of the companion, the mass of the companion would have to be
0.7(0.032/1) = 0.02 \Msun. The companion would then have to be a very
low-mass star, or a brown dwarf. Note, however, that the residuals are
relatively poorly constrained with the existing data, and that
additional observations aimed --in particular-- at confirming the
periodicity in the residuals will be needed to resolve this issue.

\section{Discussion}

\subsection{Distance to the Taurus association}

HDE~283572 was one of the few Taurus members with a parallax estimate
from Hipparcos ($\pi$ = 7.81 $\pm$ 1.30 mas; $d$ = 128$^{+26}_{-18}$
pc; Bertout et al.\ 1999). The present determination is well within
1$\sigma$ of the Hipparcos value, but more that one order of magnitude
more precise.  Bertout \& Genova (2006) estimated the distance to both
Hubble 4 ($\pi$ = 8.12 $\pm$ 1.5 mas; $d$ = 123$^{+28}_{-29}$ pc) and
HDE~283572 ($\pi$ = 7.64 $\pm$ 1.05 mas; $d$ = 131$^{+21}_{-26}$ pc)
using a modified convergent point method. Again, our values are within
1$\sigma$ of these determinations, but more than one order of
magnitude more precise. Only two other Taurus members have VLBI-based
distance determinations: T Tau ($\pi$ = 6.82 $\pm$ 0.03 mas; $d$ =
146.7 $\pm$ 0.6 pc; Loinard et al.\ 2007) and V773 Tau ($\pi$ = 6.74
$\pm$ 0.25 mas; 148.4$^{+5.7}_{-5.3}$ pc; Lestrade et al.\ 1999). The
weighted mean of these four values is $\bar{\pi}$ = 7.30 mas
($\bar{d}$ = 137.0 pc) and the r.m.s.\ dispersion about that mean 0.45
mas ($\equiv$ 9 pc). Although the number of sources with VLBI
distances remains small, we argue that the mean value represents a
good estimate of the mean distance to the Taurus association, and that
the dispersion provides a good guess of its depth. Note, however, that
the latter value was calculated as a dispersion; the corresponding
full width at half maximum (which may represent a better estimate of
the full depth of the complex) is 21 pc. In comparison, the angular
size of Taurus projected on the plane of the sky is about 10$^\circ$,
corresponding to about 23 parsecs at that distance. The significant
depth of the complex implies, in particular, that however well
measured the mean distance to Taurus may be, using it indiscriminately
for all Taurus members will result in systematic errors that may be as
large as 15\%. For higher precision, accurate individual distances to
a larger sample of Taurus members will be needed. VLBI measurements
such as those presented here most probably represent the best hope of
obtaining such a large sample in the near future.

%\begin{landscape}
\clearpage
\begin{deluxetable}{llllllll}
\tablewidth{0pt}
\rotate
\tablecaption{Space velocity for the 4 Taurus sources with VLBI-based distance determinations}
\tablehead{
\colhead{Source}       &
\colhead{}       &
\colhead{$V_r$ (km s$^{-1}$)}  &
\colhead{$V_\alpha$ (km s$^{-1}$)}  &
\colhead{$V_\delta$ (km s$^{-1}$)}  &
\colhead{$V_\ell$ (km s$^{-1}$)}  &
\colhead{$V_b$ (km s$^{-1}$)}  &
\colhead{References\tablenotemark{a}}}
\startdata
Hubble 4                    & Observed & 15.0 $\pm$ 1.7 & 2.71 $\pm$ 0.03 &  --18.2 $\pm$ 0.2  & 15.1  $\pm$ 0.1  & --10.5 $\pm$ 0.1  & 1,2 \\%
                            & Expected & 9.82 & 1.28 & --8.21 & 6.84 & --4.71\\%
HDE~283572                  & Observed & 15.0 $\pm$ 1.5 & 5.41 $\pm$ 0.04 &  --16.2 $\pm$ 0.1  & 15.55 $\pm$ 0.08 & --7.07 $\pm$ 0.05 & 1,3\\%
                            & Expected & 9.88 & 1.12 & --8.21 & 6.78 & --4.77\\%
T Tau\tablenotemark{b}      & Observed & 19.1 $\pm$ 1.2 & 8.59 $\pm$ 0.04 &  --8.90 $\pm$ 0.06 & 12.35 $\pm$ 0.05 &  +0.69 $\pm$ 0.03 & 2,4\\%
                            & Expected & 11.35 & 1.22 & --6.53 & 5.74 & --3.34\\%
V773 Tau\tablenotemark{c}   & Observed & 13.8 $\pm$ 0.9 &  0.3 $\pm$ 0.2  &  --16.4 $\pm$ 0.7  & 12.0  $\pm$ 0.5  & --11.2 $\pm$ 0.5 & 5,6\\%
                            & Expected & 9.71 & 1.48 & --8.16 & 6.89 & --4.62
\enddata
\tablenotetext{b}{1=This work; 2=Hartmann et al.\ 1986; 3=Walter et al.\ 1988; 4=Loinard et al.\ 2007; 5=Welty 1995; 6=Lestrade et al.\ 1999.}
\tablenotetext{b}{The radial velocity and proper motions used here is those of T Tau N. 
The radial velocities for T Tau Sa and T Tau Sb are available in Duch\^ene et al.\ (2002) 
and are very similar.}
\tablenotetext{c}{The radial velocity used here is that of the center of mass of the spectroscopic binary V773 Tau A+B.
\vspace{1cm}}
\end{deluxetable}
\clearpage
%\end{landscape}

\begin{figure}
\centerline{\includegraphics[width=0.65\textwidth,angle=270]{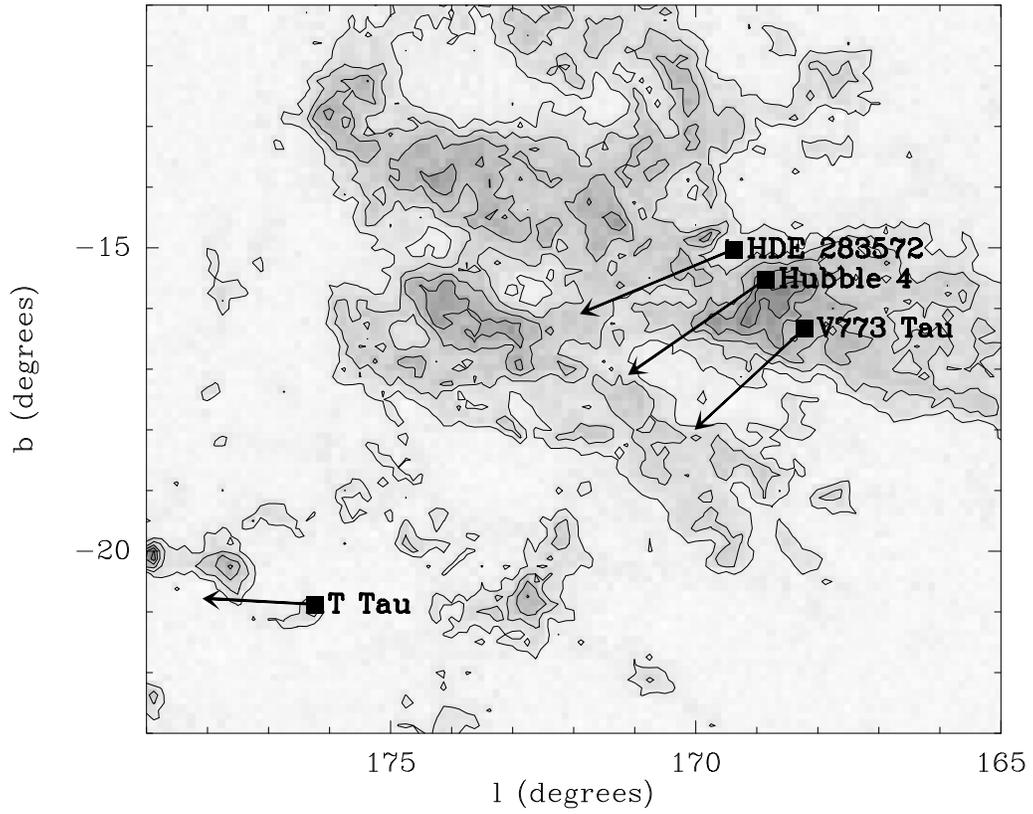}}
\caption{Positions and proper motions of the 4 sources in Taurus
with VLBI distance measurements over-imposed on the CO integrated
intensity map of Dame et al.\ (2001).}
\end{figure}

\clearpage

\subsection{Kinematics}

The tangential velocity of the four sources mentioned in the previous
section (Hubble 4, HDE~283572, T Tau and V773 Tau) can be deduced from
their measured distances and proper motions. Also, for these four
sources, radial velocities are available from the literature, so their
complete velocity vectors can be reconstructed (Tab.\ 3, Fig.\
4). Hubble 4 and HDE~283572, which are very close on the plane of the
sky and are at very similar distances also have very similar space
velocities. This strongly suggests that they belong to the same
sub-group (but see below). V773 Tau, which in projection also appears
near Hubble 4 and HDE~283572 but is at a significantly different
distance, has a somewhat different space velocity. Not surprisingly, T
Tau, at the south of the Taurus complex, has a significantly different
velocity.

The measured radial and tangential velocities can be compared to the
values expected from the differential rotation of the Galaxy. The
proper motions determined with the VLBA are measured with respect to
the Sun. To obtain the corresponding values expected theoretically, we
adopt a model for the local rotation of the Galaxy where the Oort
constants are A = 14.4 and B = --12.0 km s$^{-1}$ kpc$^{-1}$ (Allen
2000) and where the distance from the Sun to the Galactic center is
R$_\odot$ = 8.5 kpc. For the peculiar motion of the Sun (required to
transform values relative to the local standard of rest [LSR] to the
barycentric coordinates provided by the VLBA), we use U$_\odot$; =
+9.0, V$_\odot$; = +12.0, and W$_\odot$; = +7.0 km s$^{-1}$ (Allen
2000). Here, we follow the traditional convention where U runs from
the Sun to the Galactic center, V is in the Galactic plane,
perpendicular to U and positive in the direction of Galactic rotation,
and W is perpendicular to the Galactic plane, positive toward the
Galactic north pole. It is noteworthy from the comparison between the
observed and expected velocities that the Taurus members considered
here have very significant peculiar motions (of amplitude $\sim$ 10 km
s$^{-1}$). Since our measured values are very similar to the mean
radial velocities and proper motions in catalogs of optically selected
Taurus members (e.g.\ Ducourant et al.\ 2005; Bertout \& Genova 2006),
this large peculiar velocity appears to be characteristic of the
entire Taurus complex. This is a notable contrast with the stars in
the Orion cluster where the expected and measured mean proper motions
agree to better than 0.5 km s$^{-1}$ (G\'omez et al.\ 2005).

Section 4.1 and the present kinematics analysis show that if a
sufficiently large sample of Taurus members had VLBI-based distance
determinations, it would become possible to accurately map the
three-dimensional distribution of stars in the complex, as well as
their detailed kinematics. Using a dynamical analysis, it would then
become possible to estimate the total mass of the complex in a way
totally independent of the traditionally used molecular
observations. Also, coupled with pre-main sequence evolutionary models
(see below), it would become possible to study the space distribution
of stars as a function of their age, and thereby reconstruct the
history of star-formation in Taurus.

\subsection{Physical parameters of the stars}

Having measured the distance to two stars in Taurus, we are now in a
position to recalculate their luminosities, and place them better on
an isochrone. We will use here the pre-main sequence evolutionary
models of Siess et al.\ (1997) available on the World Wide Web. The
effective temperature of Hubble 4 is 4060 K (Brice\~no et al.\ 2002),
and its bolometric luminosity scaled with the present distance
determination is 2.7 (132.8/142)$^2$ = 2.4 \Lsun (Brice\~no et al.\
2002). For HDE~283572, the effective temperature is 5770 (G\"udel et
al.\ 2007) and the scaled bolometric luminosity 6.5(128.5/140)$^2$ =
5.5 \Lsun. Using these values as inputs for the evolutionary models,
we obtain $M$ = 0.7 \Msun, $R$ = 2.9 \Rsun, and $M$ = 1.6 \Msun, $R$ =
2.2 \Rsun, for Hubble 4 and HDE~283572, respectively. The
corresponding ages are 0.74 and 9.0 Myr, respectively. This last
result is quite surprising because --as mentioned earlier-- Hubble 4
and HDE~283572 are very near each other, and share the same
kinematics. In these conditions, one would expect them to be
coeval. Surprisingly, however, their ages appear to differ by one
order of magnitude

\section{Conclusions and perspectives}

In this article, we have reported multi-epoch VLBA observations of two naked
T Tauri stars in the Taurus complex, and used these data to measure their
trigonometric parallax and proper motions. Both stars appear to be
located at about 130 pc, somewhat nearer than the other two Taurus stars
(T Tauri and V773 Tau) with VLBI distance estimates (both are at $\sim$
147 pc). The declination of Hubble 4 shows small but systematic post-fit 
residuals that may be the result of an extended, time-variable magnetosphere
or of the presence of a companion, low-mass star or brown dwarf.

Hubble 4 and HDE~283572 appear to share the same kinematics, and 
most probably belong to the same Taurus sub-group. Surprisingly, however,
pre-main sequence evolutionary models suggest that their age differ
by an order of magnitude. The mean distance to Taurus obtained by averaging 
all four existing VLBI-based distance estimates is 137 pc, and the depth 
of the complex appears to be about 20 pc, very similar to the size of the 
complex projected on the plane of the sky. 

It is noteworthy that if observations similar to those presented here were
obtained for a significantly larger sample of Taurus members, it would
be possible to map the three-dimensional distribution and kinematics of the
complex, and establish the history of star-formation in this important 
nearby star-forming site. 

\acknowledgements
R.M.T., L.L. and L.F.R.\ acknowledge the financial support of DGAPA,
UNAM and CONACyT, M\'exico. We are grateful to Tom Dame for sending
us a digital version of the integrated CO(1-0) map of Taurus. The 
National Radio Astronomy Observatory is a facility of the National 
Science Foundation operated under cooperative agreement by Associated 
Universities, Inc.

\end{document}